\documentstyle[aas2pp4,psfig]{article}

\begin{document}
\title{ON IONIZATION EFFECTS AND ABUNDANCE RATIOS IN DAMPED Ly$\alpha$ SYSTEMS}
\author{Yuri I. Izotov}
\affil{Main Astronomical Observatory, Ukrainian National Academy of Sciences,
Golosiiv, Kyiv 03680, Ukraine \\ Electronic mail: izotov@mao.kiev.ua}
%
\author{Daniel Schaerer and Corinne Charbonnel}
\affil{Laboratoire d'Astrophysique, CNRS UMR 5572,
Observatoire Midi-Pyr\'en\'ees, 14, Av. E. Belin, F-31400 Toulouse, France 
\\ Electronic mail: schaerer@ast.obs-mip.fr, corinne.charbonnel@ast.obs-mip.fr}

\begin{abstract}
The similarity between observed velocity structures of Al III and singly ionized 
species in damped Ly$\alpha$ systems (DLAs) suggests the presence of ionized gas
in the regions where most metal absorption lines are formed.
A simplified model consisting of {\em Region 1)} a plane-parallel ionization 
bounded region illuminated by an internal radiation field, and {\em Region 2)} 
a neutral region with a negligible
metal content is considered. 
We calculate photoionization equilibrium models for region 1, and constrain the 
ionization parameter by the observed $N$(Al III)/$N$(Si II) column density ratio.
Under these conditions we find that ionization effects are important.

If these effects are taken into account, the element abundance ratios in DLAs are quite
consistent with those observed in Milky Way stars and in H II regions of local 
low-metallicity blue compact dwarf galaxies. In particular we cannot 
exclude the same primary N origin in both DLAs and metal-poor galaxies.
No depletion of heavy elements on dust grains needs to be invoked, although our models 
do not exclude the presence of little depletion.

Although highly simplified and relying on the strong assumption of a significantly
lower metal content in region 2, our model appears to be supported by recent data on a local DLA
and it is not in contradiction with the current knowledge on high redshift DLAs. 
If correct, it offers a clear simplification in the understanding of heavy element 
abundance ratios in DLAs and their comparison with the local Universe.

\end{abstract}


\centerline{\it\Large Accepted for publication in ApJ (scheduled in March 10, 2001,
  issue)}

\section {INTRODUCTION}

Damped Lyman $\alpha$ systems (DLAs) are the class of QSO
absorption systems showing the largest column density,
with $N$(H I) $\ga$ 2$\times$10$^{20}$ cm$^{-2}$. 
They are considered to be
the high-redshift progenitors of present-day luminous galaxies
(e.g. Wolfe 1990) and dominate the neutral gas content of the
Universe (Wolfe et al.\ 1995), a mass comparable with that of the
stars in present day spiral galaxies. This may suggest that DLAs
are the source of most of material available for star formation
at high redshift (Lanzetta et al.\ 1995). Later studies have
shown that probable optical counterparts of damped Ly$\alpha$
systems 
at $z \la 1$ are galaxies of different morphological types (Le Brun
et al.\ 1997).

Understanding the chemical evolutionary history of galaxies is
fundamental to the study of galaxy formation. Pettini et al.\
(1994, 1997) have performed extensive studies on the
metallicity of the damped Ly$\alpha$ systems
at high redshift. Their results
indicate a mean metallicity Zn/H $\approx$ 1/10 of the solar abundance
with a large dispersion. It was also noted that the 
element Cr is generally less abundant than Zn relative to their
solar proportion. Since Zn and Cr are thought to be made in
solar proportions throughout the Milky Way based on their
abundances in Galactic disk and halo stars, the overabundance
of Zn to Cr found in damped Ly$\alpha$ systems was taken as
evidence that some Cr was depleted from the gas phase by
dust grains.

The HIRES W.M.Keck telescope observations have opened the 
opportunity to study Ly$\alpha$ absorption lines
for a large number of elements including N, O, Mg, Al, Si, S, Cr, Mn,
Fe, Ni and Zn. Lu et al.\ (1996a),
Prochaska \& Wolfe (1999), Outram et al.\ (1999) and
Pettini et al.\ (2000) have found
that the relative abundance patterns 
indicate that the bulk of heavy elements in these
high-redshift galaxies
was produced by Type II supernovae;
there is little evidence for significant contribution from
low- to intermediate-mass stars or from Type Ia supernovae.

However, the current interpretation combining the effects of 
dust depletion and Type II supernovae enrichment (Lu et al.\ 1996a; 
Prochaska \& Wolfe 1999; Vladilo 1998) leads to several
inconsistencies.
In this picture the overabundance of Zn relative to Cr or Fe found in DLAs 
is attributed to selective depletion of Cr and Fe by dust grains.
However, it then turns out that 
such an interpretation is inconsistent 
with some of the other elemental abundance ratios seen in these galaxies, 
in particular N/O, S/Fe, Mn/Fe and Ti/Fe. 
This led Lu et al.\ (1996a)
and Prochaska \& Wolfe (1999) to conclude that the overabundance 
of Zn relative to Cr in damped Ly$\alpha$ galaxies may be
intrinsic to their stellar nucleosynthesis. 
They emphasize the need for another physical process to explain the
damped Ly$\alpha$ abundance patterns. 

Furthermore, comparisons of the distribution of Fe/H versus
redshift (Lu et al.\ 1996a; Prochaska \& Wolfe 2000) for the sample of 
damped Ly$\alpha$ galaxies with 
the corresponding relation for the Milky Way disk indicate
that the damped Ly$\alpha$ galaxies are much less metal-enriched 
than the Galactic disk in its past. Since there is
evidence that depletion of Fe by dust grains is relatively
unimportant, the difference in the enrichment level between
the sample of damped Ly$\alpha$ galaxies and Milky Way disk
suggests that DLAs are probably not high-redshift spiral
disks.
A similar conclusion has
been reached by Pettini et al.\ (1999). They argue that the relative
abundances of heavy elements in DLA systems are consistent with a
moderate degree of dust depletion that, once accounted for, leaves
no room for the enhancement of $\alpha$ elements over iron
seen in metal-poor stars in the Milky Way. 
Recently Centuri\'on et al. (2000) have found the S/Zn abundance 
ratio in the relatively high-metallicity DLAs to be lower than
solar ratio. This is contrary to
previous assertions that DLAs have been enriched solely by Type II
supernovae, but it can be understood if star formation
in these systems proceeded at a lower rate than in the early history
of our Galaxy (Pettini et al.\ 1999; Centuri\'on et al. 2000).

It is generally believed that the gas in DLAs is mostly
in the neutral stage due to the very high optical depth beyond the
hydrogen ionization limit (e.g., Viegas 1994; Prochaska
\& Wolfe 1996). Therefore, it is assumed that
all species with ionization potentials lower than hydrogen
are in a singly ionized stage, 
while other species are neutral.
With this assumption correction factors for ionization are small
and the ratios of column densities of ions are equal to
the ratios of element abundances. 
However, Lu et al.\ (1996a), Prochaska \& Wolfe (1999) and Wolfe \& Prochaska
(2000) find a
good correlation between the velocity structure of Al III and
singly ionized species. The ionization potential of Al$^+$
is 18.8 eV, i.e. greater than that of hydrogen. Therefore,
Al$^{+2}$ is likely present in ionized, not in neutral gas.
To explain the similarity of Al III and other low ionization
species line profiles, Howk \& Sembach (1999) and Izotov \& Thuan (1999)
proposed that these lines originate in the
same ionized region or in a mix of neutral and ionized clouds, 
and stressed the importance
of abundance correction for ionization effects. 
Indirect arguments to consider ionized gas in DLAs come from
studies of the warm ionized medium in the Milky Way and other
nearby galaxies (Howk \& Sembach 1999; Howk, Savage \& Fabian 1999;
Sembach et al. 2000; Jenkins et al. 2000).

Higher ionization species of C IV and Si IV are also observed in
the spectra of DLAs. However, in the majority of cases, the C IV 
and Si IV profiles differ from those of lower ionization species;
few cases exist where some components of the C IV and Si IV absorption 
lines coincide with e.g.\ Si II, Al II and Al III (see e.g., Lu et al.\ 1996a; 
Prochaska \& Wolfe 1999; Wolfe \& Prochaska 2000). The high ionization 
species are thus apparently formed in regions with different conditions 
and will not be considered here.
In rare cases absorption lines from neutral species with ionization potentials 
below that of hydrogen, such as C I and Mg I, are observed in DLAs.
The existence of such lines is a priori not in contradiction with the
presence of Al III, since neutral species can also be present in small
quantities in ionized gas (cf.\ Sect.\ 3).

Motivated by the close similarity between the velocity structures of
Al III and singly ionized species in DLAs, which suggests the presence
of ionized gas in the regions where these absorption lines are formed, 
we here consider a simple model that takes these effects to first order 
into  account. We examine a model of DLAs in which absorption lines of neutral 
and low-ionization heavy element species originate in ionized rather than in
neutral gas, and we explore the possible implications on 
abundance determinations in such systems.
As indicated by the presence of high ionization species and the complex
structure of the absorption line profiles seen in the high resolution spectra,  
the conditions of the absorbing gas must in reality be highly variable 
and of complex nature (see e.g.\ Lauroesch et al.\ 1996).
In the future, both more advanced analysis of the detailed line profiles and 
sub-components as well as more sophisticated models allowing
for complex geometries, various ISM phases etc.\ will be necessary to
obtain a more accurate description of the absorbing systems.
It is the hope that the present simplified model reflects the main
trends due to ionization effects.

The model assumptions are described in Section 2.
The results from the model calculations and the implications
on heavy element abundance ratios in DLAs are discussed
in Section 3. In Section 4 we briefly consider the application of 
our model to other objects and summarise 
the main results from this work.

\section{MODELS}

The above discussion shows that the properties of the damped 
Ly$\alpha$ systems
are yet poorly understood. While a significant amount of neutral 
gas is observed in these systems, it is not clear how the
regions of ionized gas are related to those of neutral
gas. What is the main source of ionization in DLAs - internal
stellar radiation or harder background intergalactic radiation?
Where are absorption lines of heavy elements formed - in ionized
or neutral gas? Is the metallicity of ionized and neutral gas
similar or not? Is heavy element depletion by dust grains
important?

In the present paper we explore the following ``multi-component'' 
picture for DLA systems: 
{\em 1)} an ionization bounded region illuminated by an internal 
radiation field complemented by {\em 2)} a neutral region with a 
lower metal content.
These working hypotheses are adopted to study the main
effects intrinsic stellar ionizing radiation may have on 
observed heavy element patterns in such conditions.

To explore the consequences of such a models we use the CLOUDY 
ionization equilibrium code (version c90.05; Ferland 1996; 
Ferland et al.\ 1998) to calculate the ionization state of region 1.
For simplicity, we assume internal stellar radiation and consider a
plane-parallel geometry. The outer boundary of the slab is determined
by the condition that a low electron temperature of 2000 K is reached.
 For the cases discussed here (low metallicity and low dust content)
this boundary condition is a good approximation for the outer boundary
of ionization-bounded H II regions.
For the ionizing spectrum, we adopt the spectrum of a typical O
star. We use an ATLAS line-blanketed model atmosphere (Kurucz 1991)
with effective temperatures $T_{\rm eff}$ = 33000 K, 40000 K and 45000K
and log $g$ = 4.0. The lower temperature is likely more appropriate
for the conditions in the interstellar medium of metal-rich spiral
galaxies and the local ISM in our Galaxy (cf.\ Sembach et al.\ 2000).
In low-metallicity environments
the temperature of the ionizing radiation tend to be higher as 
evidenced from studies of nearby extragalactic giant H II regions
(e.g.\ Stasi\'nska \& Leitherer 1996). 
The ionization parameter which is proportional
to the ratio of the number density of the ionizing photons to that of the
electrons is varied in the range
$U$ = 10$^{-5}$ -- 10$^{-2}$.
We consider the metal abundance of 0.1 solar, with relative element
abundances equivalent to those observed in the low-metallicity
BCDs (e.g., Izotov \& Thuan 1999). In particular,
C/O and N/O abundance ratios are reduced by factors of 2 and 5
respectively as compared to solar abundance ratios.
We also adopt for $^4$He the mass fraction $Y$ = 0.252 (or log He/H =
--1.07), which is higher than the primordial mass fraction $Y_{\rm p}$ = 0.245 
(Izotov et al.\ 1999), but is typical for the BCDs with the heavy element
abundance 1/10 solar (Izotov \& Thuan\ 1998). 
Dust grains with a dust-to-gas ratio of
1/10 that in Milky Way are taken into account in heating and cooling
processes, although the differences in ionic fractions derived in the models 
with and without dust grains are small.

A similar model has recently been considered by Howk \& Sembach (1999). 
The major difference with their approach is that they consider 
ionized density-bounded regions surrounding the neutral, H I--bearing 
clouds. The boundary of the ionized region is defined
where the local fraction of neutral hydrogen climbs above 10\%.
With these assumptions the fraction of neutral 
species like O$^0$, N$^0$ or Ar$^0$ is significantly reduced because they
always reside near the edges of H II region. Additionally,
the variation of the local ionization fraction of neutral hydrogen
at the edge of ionized density-bounded region results in large
variations of some low-ionization species, in particular, of Cr$^+$
and Fe$^+$. 
Further resulting differences between the model by Howk \& Sembach
(1999) and our ionization-bounded model will be discussed below.

An important feature of both models is that 
the heavy element abundances
in the ionized and neutral regions are a priori unrelated.
This implies in particular that relative abundance determinations
with respect to hydrogen, as derived from the observed column densities,
may not be meaningful or are at least uncertain.

\section{RESULTS}

\subsection{Ionization fractions of elements}

In Table 1 we show the ionization potentials and the ionization fractions 
$x$\footnote{By definition the ionization fraction $x$(X$^{+i}$) is the 
ratio of the radially integrated number density of ion X$^{+i}$ to the 
radially integrated total 
number density of the element X in all ionization stages, i.e.
$x({\rm X}^{+i}) = \int n({\rm X}^{+i}) dr / \int n({\rm X}) dr$,
the relative column density of the ion X$^{+i}$.}
of the most important ions which are observed or can potentially be observed
in DLAs, as a function of the ionization parameter $U$ for two
temperatures $T_{\rm eff}$ = 33000 K and 40000 K of ionizing radiation and the
heavy element mass fraction $Z$ = 0.1$Z_\odot$.
On the first line we also show $N$(H$^0$+H$^+$), the total hydrogen
column density across the ionized slab. Over the entire range
of $U$,
except for the highest value shown here ($ U=10^{-2}$),
$N$(H$^0$+H$^+$) remains significantly smaller than the column
density of neutral hydrogen observed in damped Ly$\alpha$ 
system\footnote{For $T_{\rm eff}$ = 40000 K and the ``typical'' value 
of $\log U \sim$ --3.25 the column density is $\sim$ 40 \% of 
the lower limit which defines damped systems. The baryonic mass
of low column density DLAs could thus be underestimated by 
up to this amount.}.
For an ionization parameter $U$ $\sim$ 10$^{-5}$ hydrogen is
essentially neutral, while at $U$ $\ga$ 10$^{-3}$ it is almostly
ionized as it is evidenced by small neutral hydrogen
fraction $x$(H$^0$) (Table 1). Calculations for lower heavy element
abundances show that the ionization fractions are not strongly changed.
Therefore, the data from Table 1 can be used for objects
with metallicity [Fe/H] $\la$ --1.

The distribution of the ionic fractions $x$(X$^i$) of selected ions
as a function of ionization parameter $U$ is shown in Figure 1. 
While the ionization fractions of all neutral and singly
ionized species decrease with increasing ionization parameter $U$,
the ionization fraction of Al$^{+2}$ is increased. Therefore, the
ratio $x$(Al$^{+2}$)/$x$(Al$^+$) = $N$(Al III) / $N$(Al II) is a very 
sensitive indicator of ionization conditions and can be used to
determine $U$. 
Unfortunately, 
although line profiles of Al II and Al III in damped Ly$\alpha$
systems are essentially similar (Lu et al.\ 1996a; Prochaska \& Wolfe
1999), in the many cases the Al II lines are saturated or marginally
saturated. This circumstance complicates direct estimates of the
ionization parameter $U$. However, very often, unsaturated lines
of Si II, Fe II and Al III are observed in the same spectra and they 
can be used to estimate $U$. For this, we assume that the observed 
heavy element abundance patterns at low metallicities are 
similar to that observed in low-metallicity halo stars, 
i.e. [Si/Fe] $\sim$ 0.3,
[Al/Fe] $\sim$ --0.4 (Timmes et al.\ 1995; Samland 1998), and adopt that
log (Al/Si)$_\odot$ $\approx$ log (Al/Fe)$_\odot$ $\approx$ --1.0 (Anders \& Grevesse 1989). 
Then, the expected
ratio of column densities 
is equal to 
\begin{equation}
\log\frac{N(\rm Al\ III)}{N(\rm Si\ II)}\sim -1.7 + \log\frac{x({\rm Al^{+2}})}
{x({\rm Si^+})}, 
\end{equation}
\begin{equation}
\log\frac{N(\rm Al\ III)}{N(\rm Fe\ II)}\sim -1.4 + \log\frac{x({\rm Al^{+2}})}
{x({\rm Fe^+})}. 
\end{equation}

The typical values of log [$N$(Al III)/$N$(Si II)] and 
log [$N$(Al III)/$N$(Fe II)] in damped Ly$\alpha$ systems
are $\sim$ --2.0 to --2.5  and $\sim$ --1.5 to --2.0, respectively, 
which translates to log [$x$(Al$^{+2}$)/$x$(Si$^+$)] $\sim$
--0.3 to --0.8, [$x$(Al$^{+2}$)/$x$(Fe$^+$)] $\sim$
--0.1 to --0.6, or log $U$ $\sim$ --3 (Fig. 1).
This estimated value of
ionization parameter is dependent on the
uncertainties in abundance determination and those for the adopted 
photoionized model and possible depletion of elements into grains.
In principle, other single ionized species, e.g.\ Zn II,  
can be used instead of Si II or Fe II. However, Si II and Fe II 
absorption features are more often
seen and measured in the DLAs, and the atomic data for these ions are known
with better precision (Howk \& Sembach 1999). 
The above estimate of $U$ is based on the assumption that heavy elements are
not present in the neutral gas surrounding the region of ionized gas.
If the neutral gas is significantly enriched in heavy elements then
an additional contribution to the column densities of Si II and Fe II comes from
neutral gas and the ionization parameter should be increased to account for
observed $N$(Al III)/$N$(Si II) and $N$(Al III)/$N$(Fe II) ratios.

Before we proceed to discuss the implications of this model when applied 
to observations of damped Ly$\alpha$ systems it is useful to recall that
some neutral species with ionization
potentials below that of hydrogen can be present in ionized gas as well.
For instance Lu et al.\ (1996a) have measured $\log N$(Mg II) $>$ 14.38 and
$\log N$(Mg I) = 12.44$\pm$0.04 in the $z_{\rm damp}$ = 0.8598 system
toward Q0454+0356 which are consistent with the predictions
of our simplified models (Table 1). Their measurements for
another $z_{\rm damp}$ = 1.2667 system toward Q0449--1326 provide only
lower limits $\log N$(Mg II) $>$ 14.1 and $\log N$(Mg I) $>$ 12.62,
also not in contradiction with the ionization model.
Prochaska \& Wolfe (1999) have derived $\log N$(C II) $>$ 14.6 and 
$\log N$(C I) $<$ 12.8 in the $z_{\rm damp}$ = 2.140 toward Q0149+33
which are consistent with predictions of ionization models.
Hence, the presence of species like C I or Mg I
does not necessarily mean that they reside in neutral gas.

A useful validity check of our model is the comparison
of the predictions of photoionization models with the observed values
of [O I/Si II]. In the case of the neutral gas model it is expected that
[O I/Si II] $\sim$ 0, while in the case of our ionization models 
[O I/Si II] = --0.5 - --0.8 depending on $T_{\rm eff}$ and for the
range log $U$ = --3.5 - --3.0 (Table 1). In the majority of cases observations of
DLA systems give only lower limits of O I abundance because of saturation
of the O I $\lambda$1302 absorption line. 
The available data in the literature shows that, in general, the lower limits 
of [O I/Si II] are consistent with values expected from the ionization
models. However, in some DLAs the lower limits of [O I/Si II]
are larger than those expected from the ionization models. Lu et al. (1996b) 
find [O I/Si II] $>$ --0.27 for the $z$ = 4.38 absorber towards 1202--0275,
while Prochaska \& Wolfe (1999) find [O I/Si II] = --0.17 in the absorber
towards Q0951--04. These values seems difficult to reconcile with 
ionization models unless a low ionization parameter is appropriate. However,
no measurements of Al III have been done for these absorbers which would allow
to constrain the ionization parameter $U$. 

Recently Molaro et al. (2000) have derived a high value of [O I/Si II] = 0.0 - 0.1
from the unsaturated O I and Si II lines in the $z$ = 3.39 absorber towards
the QSO 0000--2620. For this absorber $N$(Al III) / $N$(Si II) = --2.4
is rather high, suggesting an ionization parameter 
$\log U \sim$ --3.0 to --3.5. However, inspection of Fig.\ 1 in Prochaska
\& Wolfe (1999) indicates different profiles of Si II $\lambda$1808 and Al III
$\lambda$1854 in this absorber,
at variance with other absorbers rendering the estimate of the
ionization parameter $U$ rather uncertain. 

The above discussion suggests that large
ionized regions and abundance discontinuities might be present in some
DLAs, while other DLAs might be essentially neutral.
It is likely that DLAs may represent a mixed bag of systems with
varying ionization conditions.
While the ionization model does not pretend to explain abundance pattern in all
damped Ly$\alpha$ systems, we conclude that in many cases the derived lower 
limits of [O I/Si II] do not rule out this model.
However, we point out that better determinations of [O I/H I] are needed
to test the importance of the ionization effects in the damped Ly$\alpha$
systems. 

\subsection{Age-metallicity relation and the large spread in {\rm [Fe/H]}}

A very extensive literature exists on the interpretation of the metallicity 
of DLA as a function of redshift (e.g.\ Pettini et al.\ 1994;  Lu et al.\ 
1996a; Lindner, Fritze - von Alvensleben \& Fricke 1999; Prantzos \& 
Boissier 2000).
Among the main issues discussed are
the question of apparent evolution or non-evolution of metallicity with
$z$, the origin of the scatter, and possible observational biases affecting
this relation. A detailed comparison with such studies is beyond the scope 
of the present paper. Here, we shall only indicate the typical
metallicity shift expected in the framework of our model and mention the
main implication in terms of age-metallicity relations.

Pettini et al.\ (1994, 1999), Lu et al.\ (1996a), Prochaska \& Wolfe
(2000) noted that the chemical enrichment
processes in damped Ly$\alpha$ galaxies appear to be quite inhomogeneous. The
[Zn/H] and [Fe/H] distributions show a scatter by a factor of $\sim$ 20
with the mean value of [Zn/H] which is 0.4 larger than that of [Fe/H]
for the DLAs at $z$ $\approx$ 2 (Prochaska \& Wolfe 2000).
Furthermore, Lu et al.\ (1996a) and Pettini et al.\ (1999) 
noted that the degree of chemical enrichment
in the sample of damped Ly$\alpha$ galaxies is, on average, considerably
lower than the Milky Way disk at any given time in the past. This lead them
to conclude that the chemical enrichment of damped Ly$\alpha$ galaxies is
different from that in the disk of Milky Way. 

A possible explanation for the difference in chemical enrichment is 
simply that DLAs are in general not necessarily spiral galaxies (Le Brun 
et al.\ 1997).
Alternatively, if in the framework of our model the heavy element
lines are produced only in the ionized region (i.e.\ the neutral gas
has a very low metal content), the real Fe/H abundance in the ionized 
gas is several times higher than that derived from 
the $N$(Fe II)/$N$(H I) ratio, because of the 
non negligible upward correction for iron 
(Table 1). The upward correction for zinc is lower, and interestingly it nearly
compensates the difference
between [Fe/H] and [Zn/H] discussed by Prochaska \& Wolfe (2000). 
Then, the age-metallicity relation for damped Ly$\alpha$ galaxies is
more consistent with that for thick disk stars of our Galaxy. 
The opposite situation is of course realised if the neutral gas has the 
same metallicity as the ionized gas. 
In this case the observed Fe/H ratio is higher than real
because in ionized gas $x$(Fe$^+$) $>$ $x$(H$^0$) while in neutral gas
$x$(Fe$^+$) = $x$(H$^0$) = 1. 
As we will show below, it is, however,
difficult to adjust all observed heavy element abundance ratios
in the frame of the model with similar metallicities of neutral and ionized 
gas.

\subsection{Ionization vs Dust Depletion}

There is little doubt that some depletion of heavy elements by dust is present
in DLAs. However, the level of depletion in DLAs is 
generally smaller than that in the disk and possibly in the halo of our 
Galaxy. Pettini et al.\ (2000) adopt the model of DLA consisting only of 
neutral gas and propose for DLAs with low level of depletion
(systems with typically [Zn/Fe] $\la$ 0.3)
to correct abundances of refractory elements assuming constant
depletion factors for all these elements. 
This approach obviously leaves the question open how 
to explain observed abundances in systems with more depletion.
Besides that, the similarity of the line profiles of singly ionized ions 
and Al III and the large $N$(Al III)/$N$(Al II) ratio 
is not addressed. 

Prochaska \& Wolfe (1999) have suggested that if the overabundance of Si/Fe 
relative
to solar is indicative of dust depletion, then one might expect a correlation
between [Si/Fe] and [Zn/Fe] with the most heavily depleted regions showing
the largest Si/Fe and Zn/Fe ratios. Indeed, they found such correlation and
concluded that this appearance is consistent with that expected for dust
depletion (their Fig. 23).

Alternatively, as shown in Fig.\ 2, the regression between [Si/Fe]
and [Zn/Fe] can be explained by ionization effects. 
Dashed, solid and dotted lines in Figure
connect theoretical points [$x$(Si$^+$)/$x$(Fe$^+$), $x$(Zn$^+$)/$x$(Fe$^+$)]
in models with $Z$ = 0.1 $Z_\odot$ calculated for various $U$ values and
for three values $T_{\rm eff}$ = 33000 K, 40000 K and 45000 K of the 
effective temperature characterising the hardness of the ionizing radiation.
Note that [$x$(Si$^+$)/$x$(Fe$^+$)]$^{-1}$ and [$x$(Zn$^+$)/$x$(Fe$^+$)]$^{-1}$
are the correction factors to convert respectively Si$^+$/Fe$^+$ to Si/Fe and 
Zn$^+$/Fe$^+$ to Zn/Fe and they are independent on the intrinsic 
abundances of Si, Fe and Zn as far as these elements do not contribute
significantly to the thermal balance of the ionized gas. 
However, the 
correction factors for the considered ions can depend on the helium and
the total heavy element abundances which regulate the thermal balance and
the ionization structure. As mentioned above, the values of the ionization 
fractions $x$ are essentially metallicity independent for heavy element
mass fractions $Z \sim$ 0.1 -- 0.01 $Z_\odot$.
We point out that the quantities commonly (e.g.\ Prochaska \& Wolfe 1999)  
denoted as [Si/Fe] and [Zn/Fe] are actually
\begin{equation}
{\rm [Si/Fe]} = \log [N{\rm (Si~II)}/N{\rm (Fe~II)}] - 
\log{\rm (Si/Fe)}_\odot
\end{equation}
and
\begin{equation}
{\rm [Zn/Fe]} = \log [N{\rm (Zn~II)}/N{\rm (Fe~II)}] - 
\log{\rm (Zn/Fe)}_\odot,
\end{equation}
respectively. The total abundance ratios are then defined as
\begin{equation}
{\rm [Si/Fe]_{cor}} = {\rm [Si/Fe]} - \log[x({\rm Si}^+)/x({\rm Fe}^+)]
\end{equation}
and
\begin{equation}
{\rm [Zn/Fe]_{cor}} = {\rm [Zn/Fe]} - \log[x({\rm Zn}^+)/x({\rm Fe}^+)].
\end{equation}

We find that both the observed range of $x$(Al$^{+2}$) / $x$(Si$^+$)
and [Zn/Fe] are consistent with same ionization parameter for the 
ionizing radiation with $T_{\rm eff}$ = 33000K and 40000K.
Figure 2 shows that if one applies the correction for ionization effects 
for the radiation effective temperature of 40000 K,
assuming that after correction [Zn/Fe]$_{\rm cor}$ = 0, the resulting 
[Si/Fe]$_{\rm cor}$ value 
remains on average above zero with a small spread of points (see also 
Fig.\ 4), and no additional correction 
for dust depletion is required. However, models with harder radiation
with $T_{\rm eff}$ = 45000 K do not reproduce the observed trend. 
If our approach is
correct then we obtain important constraint on the hardness
of the ionizing stellar radiation in DLAs from Figure 2.
Note that with the hard intergalactic background radiation field 
considered by Howk \& Sembach (1999) the observed [Zn/Fe] vs [Si/Fe] relation
cannot be reproduced because of their adopted boundary conditions.
In this case (see their Fig. 1) the correction for ionization 
of [Zn/Fe] is positive, while that of [Si/Fe] is negative. 
Even if an ionization bounded model with the hard intergalactic 
background radiation field is considered, the correction for ionization
of [Zn/Fe], though smaller, remains positive.
In our case both corrections are negative.

In summary, it is possible to reproduce the observed $N$(Al III)/$N$(Si II)
ratios and the trend seen in [Si/Fe] vs [Zn/Fe] diagram assuming ionization 
by a moderately hard stellar ionizing radiation field. 
In principle, no correction
for dust depletion needs to be invoked, although it might be present
in small amounts and can easily be taken into account in the
ionization model without significant changing the main features of 
the [Si/Fe] vs [Zn/Fe] distribution. 

Additional support of the ionization model comes from Fig. 3 
where we show the dependence of [Al III/Fe II] vs [Zn II/Fe II]. Models
with $T_{\rm eff}$ = 33000K (dashed line) and 40000K (solid line) fit quite well
the observed points from Prochaska \& Wolfe (1999) and Lu et al. (1996a)
at the values of ionization parameter $U$ close to that obtained from 
Fig. 2.

\subsection{Abundance ratios in ionization bounded DLAs}

The above conclusion is broadly consistent with 
expectations for other heavy element abundance ratios. 
To illustrate this we apply a ``typical'' ionization correction factor to the
observational data. To do so we 
adopt the characteristic correction factors of the model with gas ionized
by the radiation with $T_{\rm eff}$ = 40000 K. We consider 
the metallicity $Z$ = 0.1
$Z_\odot$ as representative for the sample of DLA systems and adopt log $U$ =
--3.25.
Then, our calculations yield the following mean ionization correction factors 
$\Delta \log[{\rm X/H}]$ defined by 
$ \log[{\rm X/H}]_{\rm corrected} = \log[{\rm X/H}]_{\rm observed} 
  + \Delta \log[{\rm X/H}]$:
$\Delta$log[Al/Fe] = --0.63, $\Delta$log[Si/Fe] = --0.31, 
$\Delta$log[S/Fe] = --0.32,
$\Delta$log[Ti/Fe] = --0.12, $\Delta$log[Cr/Fe] = +0.16, 
$\Delta$log[Mn/Fe] = --0.01, 
$\Delta$log[Ni/Fe] = --0.38, $\Delta$log[Zn/Fe] = --0.40.
Note the small corrections for [Ti/Fe] and [Mn/Fe] which remain
almost unchanged while the downward correction of [Zn/Fe] is high.
Interestingly, it is the same as the difference between mean [Fe/H] and [Zn/H] values
for DLAs at $z$ $\approx$ 2 from Prochaska \& Wolfe (2000).
The case of nitrogen is discussed separately below.

The above ``typical'' correction factors have been applied to the
sample of damped Ly$\alpha$ galaxies from Lu et al.\ (1996a), Prochaska \&
Wolfe (1999) and Pettini et al.\ (2000). 
Only galaxies with definitely derived column densities are 
included and we do not consider objects with lower or upper limits. 
The original (open symbols) and corrected (filled symbols) abundance 
ratios are shown in Fig. 4.
As pointed out earlier, an upward correction of [Fe/H] must also be applied
because of the large column density of neutral hydrogen. However,
this correction is uncertain and does not change the relative 
heavy element abundances.
Therefore, we simply use [Fe/H] values derived by Lu et al.\ (1996a), 
Prochaska \& Wolfe (1999) and Pettini et al.\ (2000) 
keeping in mind that correction might be up to of the order of +1 dex.
For 4 DLAs of the sample the [Fe/H] are not available. We adopt
[Fe/H] = [Cr/H] for those systems.
For [Ni/Fe] we have applied an additional upward correction 
which comes from the new measurements of oscillator strengths for 
Ni II from UV absorption lines (Fedchak \& Lawler 1999);
older $f$-values were used by Prochaska \& Wolfe (1999) and Lu et al.\ 
(1996a). Although the correction is different for different Ni II transitions,
its mean value is $\sim$ 0.3 dex, which we adopt for the sake of
simplicity. 
We now discuss the behaviour of the element ratios of 
Si, Zn, S, Mn, Cr, Ni, Ti, and Al with respect to Fe.

{\bf Silicon, Zinc:} 
The case of Si and Zn has already been treated before (Sect.\ 3.3).
In short, it can be seen from Fig.\ 4 that the corrected [Zn/Fe] values 
scatter around zero, while [Si/Fe] remains positive
suggesting the production of this element in Type II supernovae.

{\bf Sulphur:} 
[S/Fe] shows the same SNII like pattern as [Si/Fe].
We note that if the gas in DLA galaxies is ionized by softer 
stellar radiation with $T_{\rm eff}$ = 33000 K, the downward correction of 
S/Fe abundance ratio is 0.17 dex larger and in this case [S/Fe] $<$ 0, in 
contradiction with the production of S in Type II supernovae. This 
difference in S correction factors
also favours harder ionization radiation for the low-metallicity DLAs as 
compared to that in the Milky Way.
A new analysis of the S/Zn abundance ratio in a sample of DLAs was recently
presented by  Centuri\'on et al. (2000).
They find [S/Zn] $<$ 0 in DLAs with [Zn/H] $\ga$ --1.0.
Both S and Zn are not depleted into the dust grains and the correction
for dust depletion does not change S/Zn ratio. Centuri\'on et al. (2000)
thus conclude that such unusual abundance ratios suggest that the DLA 
galaxies are objects with low star formation rates such as low surface
brightness galaxies or dwarf galaxies. 
Based on our model, the correction for the ionization results in the increase
of both [S/Zn] (Table 1) and [Zn/H] and in better agreement
with [S/Fe] for halo stars. If instead the model with the hard background
intergalactic radiation is adopted, the correction of [S/Zn] is negative
and it increases the difference between [S/Zn] in DLAs and [S/Fe] in halo
stars.

{\bf Manganese, Chromium:}
[Mn/Fe] remains nearly unchanged because of small correction factor
for this ratio. The Cr/Fe ratio is overabundant by $\sim$ 0.2 dex with respect
to the solar value, which seems to be in contradiction with the stellar data.
This may constitute a difficulty for our model. Uncertainties in oscillator 
strengths for Cr II transitions are better than 10\% (Bergeson \& Lawler
(1993). However uncertainties in the ionization and recombination 
cross-sections should be examined. 
Adopting a higher effective temperature decreases the overabundance of Cr 
relative to Fe, as can be seen from Table 1.
The Mn/Fe ratio remains
underabundant supporting the origin of this odd-element in Type II supernovae.

{\bf Nickel:}
Note that [Ni/Fe] calculated with the new oscillator strengths before the 
ionization correction (open circles) are already above zero. This poses a 
problem in the explanation of [Ni/Fe] when dust depletion is taken into 
account, since this correction shifts the points upward.
Although a large negative correction of --0.38 is predicted for [Ni/Fe]
in the ionization model, it
is almostly compensated by the upward shift of points introduced by the new 
oscillator strengths by Fedchak \& Lawler (1999). The points in [Ni/Fe] vs
[Fe/H] diagram scatter around the mean [Ni/Fe] value of --0.2.

{\bf Titanium:}
Prochaska \& Wolfe (1999) pointed out problems in explaining the Ti/Fe 
abundance ratio in the ``dust depletion'' model, since Ti is expected to be
more heavily incorporated into dust grains. Hence, the correction for
depletion shifts observed [Ti/Fe] to larger values. In our model, the 
correction of [Ti/Fe] for ionization effects is negative and small. 
Therefore, its application to Prochaska \& Wolfe (1999) data leaves Ti 
moderately overabundant relative to iron (filled squares in Fig. 4), in 
agreement with that observed in galactic halo stars. 

{\bf Aluminium:}
Our model predicts the largest downward correction factor of --0.63 dex
for Al/Fe abundance ratio. In the Prochaska \& Wolfe (1999) sample the [Al/Fe]
scatter around zero value. This is likely at variance with what is observed in
lowest-metallicity halo stars where [Al/Fe] scatters below zero with mean value
of $\sim$ --0.4 (Ryan et al.\ 1996; Timmes et al.\ 1995;
Samland 1998). Our correction eliminates this difference and suggests that
the odd-element Al seen in damped Ly$\alpha$ galaxies is likely also produced
by Type II supernovae. If instead the hard intergalactic ionization 
radiation is applied, the correction of [Al/Fe] is $\sim$ +0.1 and shifts 
points upward in the direction opposite to the nucleosynthetic predictions.

Overall, we find that in the framework of our simplified model of 
photoionized gas and despite the uncertainties in the atomic data, the
abundances in DLA systems reproduce quite well observed halo star abundance 
patterns and lead to a consistent picture.
Several problems pertaining to neutral DLA models are thus eliminated.
In the new model, there is no need to
take depletion of heavy elements into dust grains into account,
although small depletion of heavy elements might be present.

In the framework of the proposed model the abundance ratios of neutral 
and singly ionized species N I / O I,  Si II/ S II, Mn II / Fe II (Table 1)
depend only weakly on ionization effects and can be used as a good
measure of the total abundance ratios.

\subsection{On the origin of nitrogen}

We now turn to a discussion of the nitrogen abundance, which is of 
particular interest since it is one of the elements observable
both in metal-poor HII regions and DLAs, and because its primary 
and/or secondary origin remains debated (see e.g.\ Henry, Edmunds 
\& K\"oppen 2000).

The heavy element abundance patterns seen in DLA galaxies show several
similarities with those found in local blue compact dwarf galaxies.
Izotov \& Thuan (1999), based on the large sample of low-metallicity BCDs,
have demonstrated that all elements of $\alpha$ process are overabundant
relative to iron and hence they are produced in Type II supernovae.

In this respect
the N/O ratios seem, at first glance, to constitute the major difference. The 
N/O ratios in damped Ly$\alpha$ systems appear to be significantly lower than 
those measured in low-metallicity BCDs (cf.\ Fig. 5).
For the most metal deficient of these objects with [O/H] $\la$ --1.3 
Thuan, Izotov \& Lipovetsky (1995) and
Izotov \& Thuan (1999) have derived a mean value of log N/O = --1.60 with a 
very small dispersion. They argue that this value constitutes 
a lower limit to log N/O in BCDs, which is set by primary N production in 
massive stars.
Pettini, Lipman \& Hunstead (1995) have found an upper limit log N/O 
$\la$ --1.9 in one damped Ly$\alpha$ system, 0.3 dex below the value of 
Izotov \& Thuan. However, within the errors the log N/O value derived by
Pettini et al.\ (1995) is still consistent with the mean value for BCDs.
Furthermore, Lu, Sargent \&
Barlow (1998) have compiled N/Si abundance ratios in 15 damped Ly$\alpha$
galaxies, and have set upper limits for log N/O ranging from from --1.2 to 
--2.1.
Whereas all these measurements constitute upper limits,
an actual value of [N/Si] = --1.70 has been derived in one system, 1946+7658, 
by Lu et al.\ (1998).
From these DLA measurements of N/O and N/Si, Pettini et al.\ (1995) conclude 
that both primary and secondary nitrogen production are important over the 
whole range of metallicity measured in damped Ly$\alpha$ systems,
while Lu et al.\ (1998) favor the time-delayed primary nitrogen production
by intermediate-mass stars.

Can these different scenarios be reconciled ?
How well are these differences established, especially in view of the
possible ionization corrections discussed above for other elements detected
in DLA systems ? 
Indeed if our model holds, ionization effects are also important for the 
N/Si abundance ratio. 
The inspection of Table 1 and Figure 1 shows that correction factors for the 
observed $N$(N$^0$)/$N$(Si$^+$) ratio 
increase with increasing $U$. For the typical value log $U$ = --3.25
adopted for the above analysis of other heavy element abundance ratios 
(cf.\ Fig.\ 4) we find $\Delta$log[$x$(N$^0$)/$x$(Si$^+$)] = 
0.72.\footnote{In the models of Howk \& Sembach (1999) the correction 
$\Delta$log[$x$(N$^0$)/$x$(Si$^+$)] is significantly larger (1.48 dex)
because of the assumption of density-bounded H II
region in their case.}

The observations of N/O in BCDs (Izotov \& Thuan 1999, filled circles) and the 
DLA data from Lu et al.\ (1998) and Outram et al.\ (1999) (open circles) 
assuming neutral systems are plotted in Figure 5. 
Overplotted for illustration is the contribution of primary and secondary 
nitrogen production as well as the combined production by both mechanisms.
The ``typical'' ionization correction for N/Si and Si/H\footnote{Only the
correction for Si/H in ionized region is taken into account. Larger 
upward corrections for Si/H should be applied if neutral hydrogen region is
considered.}
is shown by the arrow in the lower left corner of Figure 5.
Taking into account such a correction, the vast majority of measurements for 
DLAs are found at or above the N/O value from the BCDs.
This suggests that a significant part of the difference in [N/Si] can be 
explained by ionization effects.
The typical correction brings even the lowest value of [N/Si] $\sim$ --1.7, 
measured in the $z_{\rm damp}$ = 2.8448 damped Ly$\alpha$ system toward 
Q1946 + 7658, to [N/Si] $\sim$ --0.98 after correction for ionization, 
fairly close to the [N/O] values for the BCDs.

The open square in Fig.\ 5 shows the N/O abundance ratio derived recently by Molaro et
al. (2000) in a DLA with $z_{\rm abs}$ = 3.3901 from the unsaturated O I and N I
absorption lines. Potentially, the N/O abundance ratio derived directly from N I 
and O I (instead of using Si as a proxy) is the least model-dependent.
Also it has the advantage of being very little dependent on the correction for 
ionization (see Table 1).
Most interestingly, the N/O value of Molaro et al.\ (2000) is in very good 
agreement with N/O derived in the low-metallicity BCDs.

In any case, if the typical ionization corrections derived from our model apply, 
we have to conclude
that the bulk of the present data does not show a significant difference in N/O 
abundance ratios between low-metallicity BCDs and DLA systems implying
similar origin of these two elements.
We shall now briefly summarise recent progress in stellar models
regarding the nucleosynthetic origin of nitrogen.

Interestingly both intermediate-mass and
massive stars are found to be potential producers of primary $^{14}$N
in the most recent stellar models.
Intermediate mass stars ($M_{\rm initial} \ga$ 3.5 $M_{\odot}$) 
produce primary nitrogen when the CN cycle operates in their convective 
envelope during the AGB phase (e.g. Renzini \& Voli 1981). 
This production results from the combined effect of the 3rd dredge-up
and envelope burning, and thus varies both with stellar mass
and metallicity (e.g.\ Forestini \& Charbonnel (1997) and references therein). 
It is also highly dependent on the parameters which intervene in the description 
of AGB stars. Among them, the mixing length and the dredge-up parameters which are 
empirically calibrated to reproduce peculiar observed properties of different
AGB populations (in particular the observed luminosity function of carbon stars in
the LMC and SMC), and which influence the efficiency of the burning inside the
envelope. In addition, the mass loss prescription which remains uncertain 
(e.g. Vassiliadis \& Wood 1993) has a crucial impact on the theoretical 
nucleosynthesis by controlling the number of thermal pulses undergone by the 
AGB star and the duration of the envelope burning. 
The most recent models find a $^{14}$N production similar to that of
Renzini \& Voli (1981, $\alpha=1.5$ case with the largest N production)
at [O/H] $\sim$ --0.7 but which decreases with increasing metallicity
(see Marigo 1998; Marigo et al.\ 1998; Lattanzio, Forestini \& Charbonnel 1999).
Intermediate mass stars are thus potential primary $^{14}$N producers.
However, no calculations are yet available in the metallicity range of interest
for our study. The behaviour at  [O/H] $<$ --0.7 remains to be investigated. 

Regarding massive stars it is well known that standard stellar evolution 
models do not produce primary nitrogen (e.g.\ Woosley \& Weaver 1995).
However, considerable advances have been made recently to include rotation and its 
effects on the transport of elements and angular momentum in the description of
stellar interiors and evolution (see review of Maeder \& Meynet 2000).
The corresponding new models for massive stars, which solve a number of difficulties
encountered by ``standard'' models (cf.\ Maeder 1995), also produce
primary nitrogen (Maeder 1999; Langer et al.\ 1999).
In particular the models of Maeder (1999) reproduce well the N excess 
in SMC supergiants found by Venn (1999), and suggest that these stars
could be responsible for primary N production at early epochs. 
Increasing rotational velocities found at low O/H (Maeder, Grebel \& Mermilliod
1999) could favor the primary N production in metal-poor environments.
Quantitative yields from these new stellar models have still to be awaited.

Last, but not least, such new yields from intermediate and massive stars will 
be included in chemical evolution models to fully exploit all the observational
constraints regarding N and other elements. This should hopefully solve
the question on the origin of primary N.
Independently of the outcome, the conclusion from our work is
that in the light of possible ionization corrections for DLAs
one cannot presently exclude the same primary origin of N in both DLAs and 
nearby metal-poor galaxies.

\section {DISCUSSION AND SUMMARY}

The above analysis shows the importance of consideration of the ionization
effects in the DLAs element abundance studies which was pointed out earlier
by Prochaska \& Wolfe (1999), Howk \& Sembach (1999) and 
Izotov \& Thuan (1999). However, at variance with many previous studies 
assuming that the heavy element absorption lines in DLA galaxies are formed in
the neutral gas, we consider the following simplified ``multi-component'' 
picture for DLA systems: {\em Region 1)} A plane-parallel ionization bounded 
region illuminated by an internal radiation field complemented by 
{\em Region 2)}, a neutral region with a lower metal content. 

Interestingly it appears that the structure of at least some DLA systems 
might be similar to that observed in nearby low-metallicity blue
compact dwarf galaxies. For example, the BCD galaxy SBS 0335--052
showing supergiant H II regions over a region of $\sim$ 7 kpc
with ongoing star formation embedded in much larger clouds of neutral 
gas with a column density $N$(H I) as high as 7$\times$10$^{21}$ cm$^{-2}$,
can be considered as a local damped Ly$\alpha$ system (e.g., Thuan \& 
Izotov 1997).

Another interesting parallel is the most metal-poor BCD galaxy 
I Zw 18, which shows a high hydrogen column density 
$N$(H I) $\approx$ 3.5$\times$10$^{21}$ cm$^{-2}$ (Kunth et al.\ 1994)
thus also qualifying as a local DLA system, and for which recent 
{\sl FUSE} spectroscopic observations were obtained by Vidal-Madjar et al.\ 
(2000).
In contrast with high-redshift DLAs the structure of I Zw 18 is known 
better. It consists of two central regions of active star formation ionizing 
the ambient gas. 
Radio observations and
the presence of strong Ly$\alpha$ absorption line
indicate that the H II region is embedded in the outer H I cloud. 
Measurements of the column densities of N I, Si II, Ar I and Fe II derived 
from the {\sl FUSE} spectrum of I Zw 18 were kindly made available to us
by Alfred Vidal-Madjar (2000, private communication).
It is striking that the column density ratios given in Table 2 (column 2) 
are very different from those observed in the H II regions (NW and SE 
components) of I Zw 18 and other low-metallicity BCDs also given in the 
Table 2. We suspect that such large differences are due to ionization effects. 
To analyse the influence of the ionization on the observed column densities
we consider a spherically-symmetric ionization-bounded H II region model with
the parameters derived from the observations (see e.g.\ Hunter \& Thronson 1995; 
De Mello et al.\ 1998). In particular, the heavy element mass fraction
in H II region is adopted to be 1/50 $Z_\odot$, while for the neutral gas
we assume zero metallicity. 

The results of the calculations are shown in Table 2, where the observed
ratio from {\sl FUSE} (col.\ 2) and the ionization corrected column density 
ratio (col.\ 3) are given.
Note the good agreement with the abundance ratios derived from the 
{\sl HST} and ground-based observations of I Zw 18 and other BCDs, while
solar ratios do not match corrected ratios. 
To obtain the total hydrogen column density a large amount of gas must now
be added to that of the ionized gas (cf.\ Table 2).
However, since the heavy element ratios are already close to the observed
values from the H II regions in BCDs, there is little room for additional
metals in the neutral phase. 
Hence, this analysis provides some support in favor of the basic 
assumption of our model that neutral gas in DLAs systems might be more 
metal-poor as compared to the ionized gas. 

From our simplified model (see Sect.\ 2) the following main results
are obtained (Sect.\ 3):

1. Using the observed $N$(Al III)/$N$(Si II) column density ratio to constrain
the ionization parameter $U$, one obtains almost complete ionization of 
hydrogen in regions with a column density $N$(H$^0$+H$^+$) $\sim$ 10$^{20}$ cm$^{-2}$ 
(region 1), several times smaller than the observed HI column in DLA systems.
In the ionized region Fe is mainly in the Fe$^{+2}$ form.
Both effects imply that the true Fe/H abundance in the ionized gas must be
larger (by up to one order of magnitude) than that derived from the 
$N$(Fe II)/$N$(H I) ratio.
Compared to [Fe/H] derived from a neutral DLA model, such ionized systems have 
an age-metallicity relation in better agreement with that of the thick disk of 
the Milky Way.
However, in general, due to the assumption of a significant metal-deficiency in
the neutral region (2), abundance ratios with respect to hydrogen are a priori
not meaningful when the total absorbing column is taken into account.

2. Accounting for ionization effects improves the 
agreement between DLAs and Milky Way stellar thick disk measurements.
Although not excluded, no significant depletion of elements by dust grains is 
required. The majority of heavy element abundance ratios are brought into 
agreement with those expected from the production by Type II supernovae.
The apparent difference between [N/O] derived in local blue
compact dwarf galaxies and [N/Si] derived in DLA systems can be fairly well
explained by ionization effects in the latter. 
Thus one cannot currently exclude the same primary N origin in both 
DLAs and metal-poor galaxies.

The above findings lead to a more consistent picture for the heavy
element pattern in DLAs, and render our simple model attractive.

The main sources of uncertainties potentially affecting our results are:
poor atomic data (oscillator strengths, recombination coefficients etc.\ for 
the photoionization models, affecting e.g.\  Zn, Cr, and Ni),
the unknown structure and geometry of the absorbing systems,
the spatial distribution of ionizing sources and their radiation field,
and uncertainties related to the measurement of column densities.
Future investigations should hopefully be able to improve on
these issues and to assess the validity of our picture.
For example, quantitative studies of the line profile structure of 
different ions, and detailed ionization properties of individual systems and 
large samples, should be extremely useful.
It is the hope that the proposed simplified model reflects the main
trends due to ionization effects in DLAs. If correct, our picture of 
DLA systems partly ionized by internal radiation offers a clear 
simplification in the understanding of heavy element abundance ratios 
in DLAs and their comparison with the local Universe.

\acknowledgements
Y.I.I. thanks the staff of the Observatoire Midi-Pyr\'en\'ees
for their kind hospitality, and acknowledges an invitation by the 
Universit\'e Paul Sabatier in Toulouse. 
Alfred Vidal-Madjar kindly provided us with results of {\sl FUSE}
observations.
We thank Richard F. Green and an anonymous referee for useful
comments on this paper.
This international collaboration was possible thanks to the  
partial financial support of INTAS grant No. 97-0033
for which Y.I.I. and D.S. are grateful. Y.I.I. acknowledges the partial 
financial support of the Joint Research Project between Eastern Europe and
Switzerland (SCOPE) No. 7UKPJ62178.


\clearpage

%
%
\begin{figure}
\plotfiddle{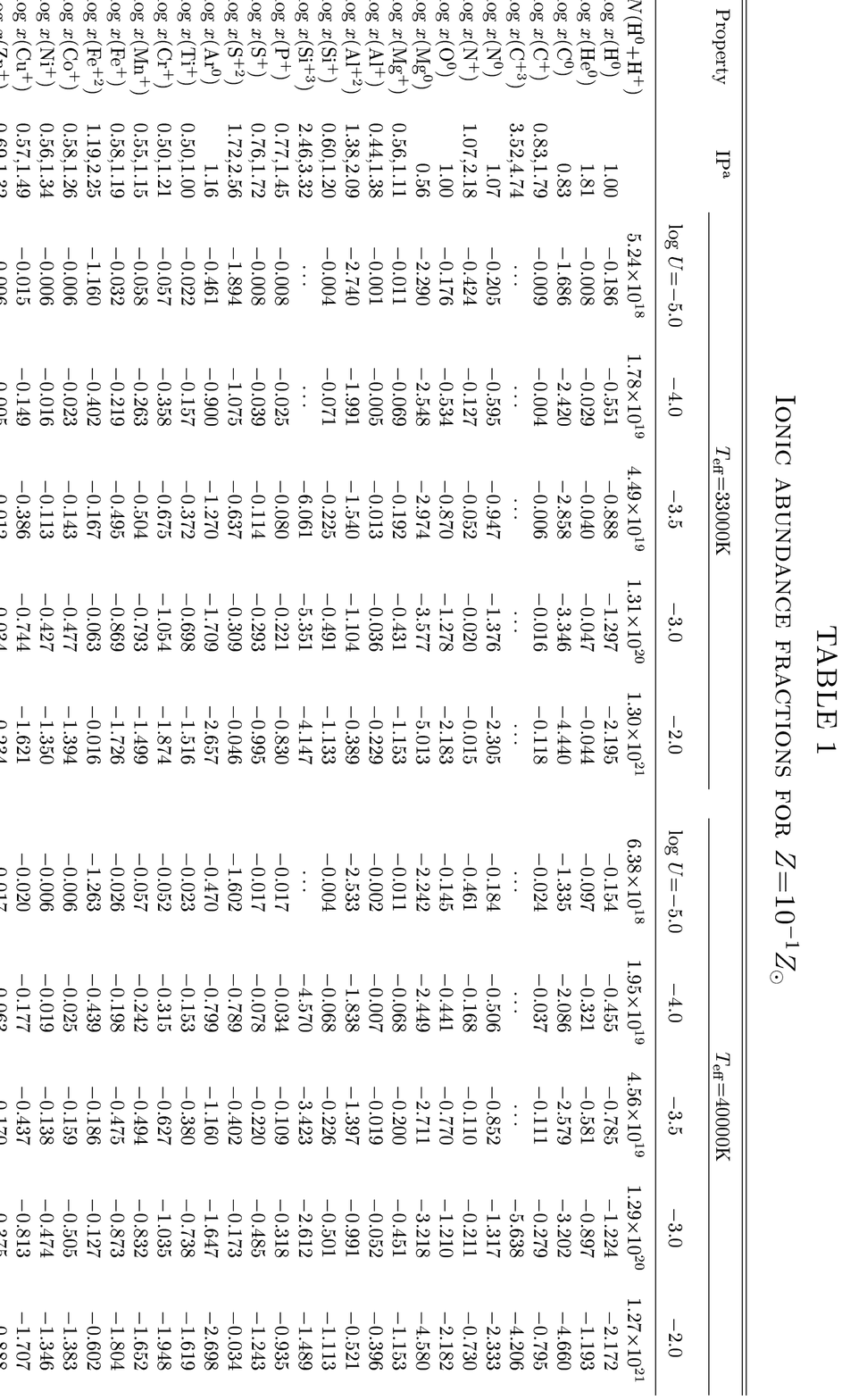}{0.cm}{180.}{80.}{80.}{300.}{380.}
\end{figure}

\clearpage
\begin{deluxetable}{lcccccc}
\tablenum{2}
\tablecolumns{6}
\tablewidth{470pt}
\tablecaption{Abundance ratios in I Zw 18 from {\sl FUSE}, {\sl HST} 
and ground-based observations}
\tablehead{
\colhead{}&\colhead{{\sl FUSE}\tablenotemark{a}}&\colhead{Model}&
  \colhead{I Zw 18 (NW)\tablenotemark{b}}&\colhead{I Zw 18 (SE)\tablenotemark{b}}&
  \colhead{BCDs\tablenotemark{c}}&\colhead{Sun\tablenotemark{d}}
 }
\startdata
$N$(H), cm$^{-2}$   &2.1$\times$10$^{21}$\tablenotemark{e}&7.7$\times$10$^{20}\tablenotemark{f}$&\nodata&\nodata&\nodata&\nodata \nl
$D$, Mpc\tablenotemark{g}     &\nodata&10&\nodata&\nodata&\nodata&\nodata \nl
log $Q$, s$^{-1}$\tablenotemark{h}     &\nodata&51.6&\nodata&\nodata&\nodata&\nodata \nl
log $R_{\rm in}$, cm\tablenotemark{i}&\nodata&20.08&\nodata&\nodata&\nodata&\nodata \nl
$T_{\rm eff}$, K\tablenotemark{j}&\nodata&50000&\nodata&\nodata&\nodata&\nodata \nl
$N$, cm$^{-3}$\tablenotemark{k}&\nodata&100&\nodata&\nodata&\nodata&\nodata \nl
$f$\tablenotemark{l}&\nodata&0.1&\nodata&\nodata&\nodata&\nodata \nl
log N/Si\tablenotemark{m}    &--0.77&--0.14&\nodata&--0.09$\pm$0.23&--0.10$\pm$0.06&$+0.32$ \nl
log Ar/Si\tablenotemark{m}   &--1.32&--0.57&--0.61$\pm$0.22&--0.74$\pm$0.22&--0.72$\pm$0.09&$-0.95$ \nl
log Fe/Si\tablenotemark{m}   &--0.77&--0.42&\nodata&--0.32$\pm$0.24&--0.24$\pm$0.12&$-0.04$ \nl
\enddata
\tablenotetext{a}{Vidal-Madjar et al. (2000), Vidal-Madjar (2000, private communication).}
\tablenotetext{b}{Si and Fe abundances are from Izotov \& Thuan (1999), abundances
of other elements are from Izotov et al. (1999).}
\tablenotetext{c}{Mean abundance ratios for the BCDs with oxygen abundance
12 + log O/H $\leq$ 7.60 (Izotov \& Thuan 1999).}
\tablenotetext{d}{Anders \& Grevesse (1989).}
\tablenotetext{e}{Neutral hydrogen column number density.}
\tablenotetext{f}{Ionized hydrogen column number density.}
\tablenotetext{g}{The distance to I Zw 18.}
\tablenotetext{h}{Total number of the ionizing photons.}
\tablenotetext{i}{Inner radius of the ionized shell.}
\tablenotetext{j}{Effective temperature of the ionizing radiation.}
\tablenotetext{k}{Number density in H II region.}
\tablenotetext{l}{Volume filling factor.}
\tablenotetext{m}{Second column: column density ratios
$N$(NI)/$N$(SiII), $N$(ArI)/$N$(SiII), $N$(FeII)/$N$(SiII) (dex) from
the {\sl FUSE} observations. Other columns: total abundance ratios (dex).
%
}

\end{deluxetable}

\clearpage
%
%
\begin{figure*}
\centerline{\psfig{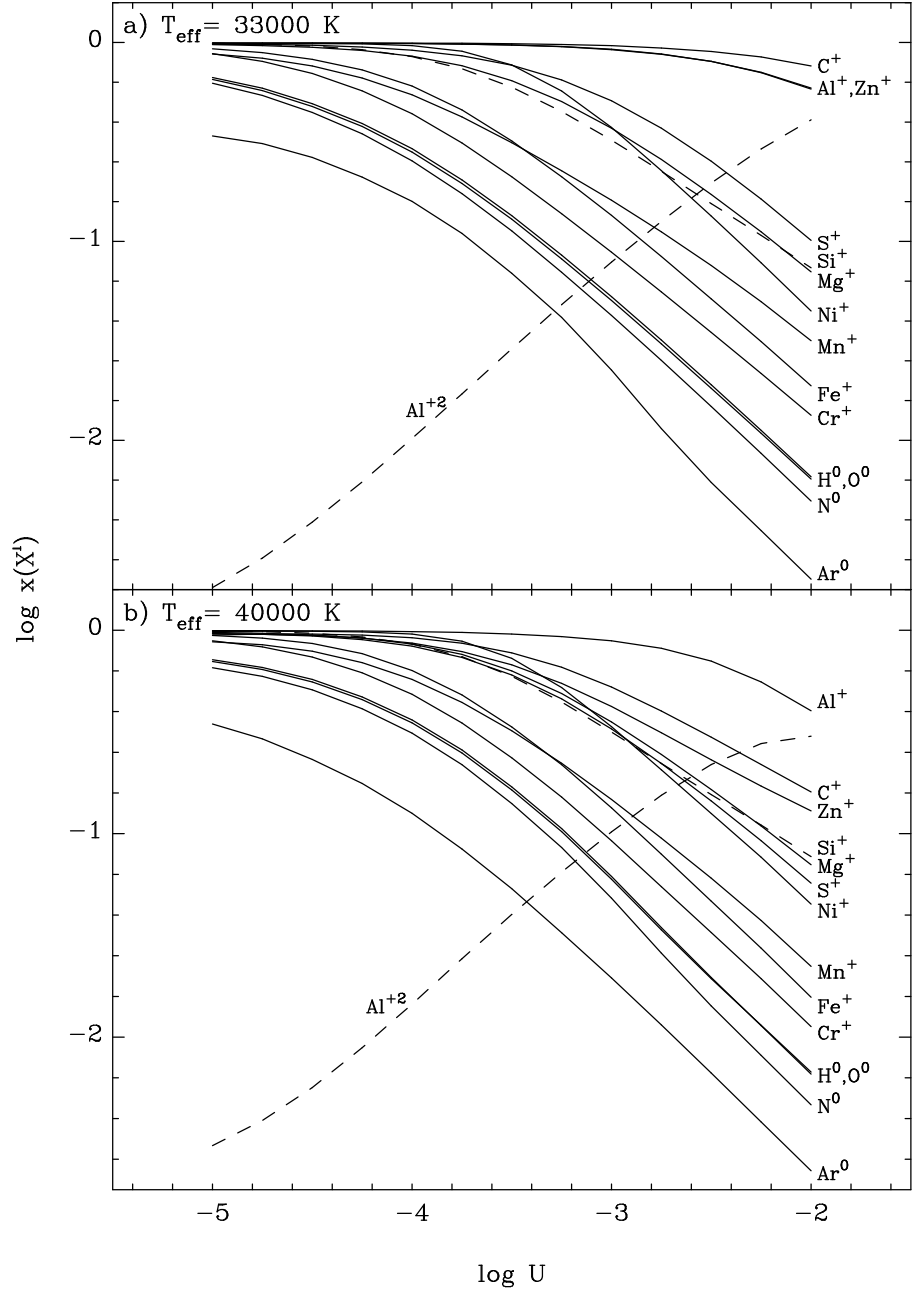}}
\caption{ Ionic fractions of elements along the line of sight
$x$(X$^i$) as a function of ionization parameter $U$ for the closed 
plane-parallel H II region models ionized by stellar radiation with (a)
$T_{\rm eff}$ = 33000 K and (b) 40000 K. The curves for Al$^{+2}$
and Si$^{+}$ are shown by dashed lines.}
\end{figure*}

%
%
\begin{figure*}
\centerline{\psfig{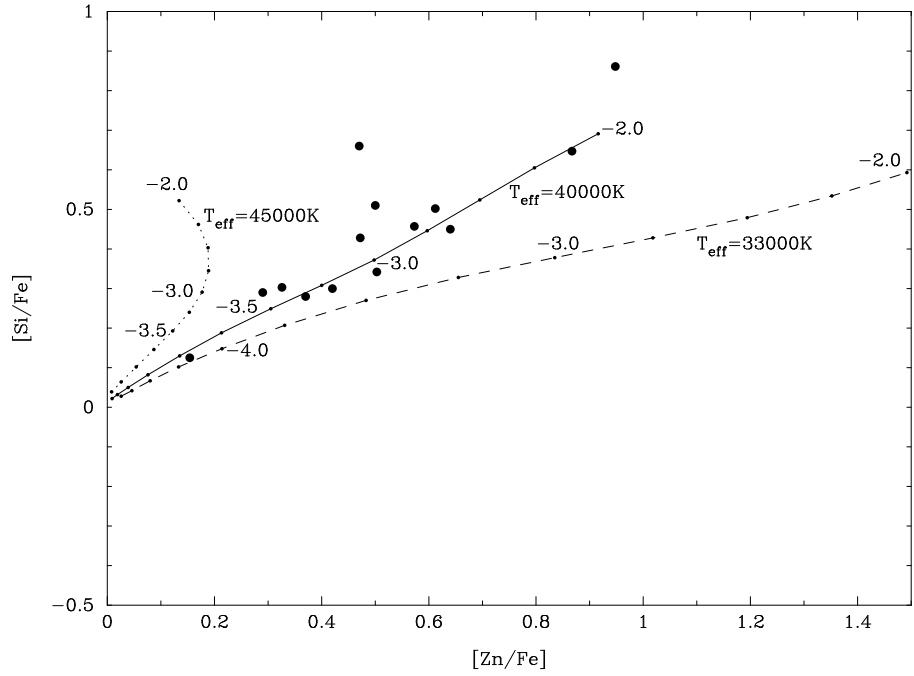}}
\caption{The [Si/Fe] vs [Zn/Fe] relation. 
The filled circles are
observational points from Prochaska \& Wolfe (1999). Dashed, solid and dotted 
lines 
connect theoretical points [$x$(Si$^+$)/$x$(Fe$^+$), $x$(Zn$^+$)/$x$(Fe$^+$)]
in models with $Z$ = 0.1 $Z_\odot$ for varying values of the ionization 
parameter $U$ (in steps of 0.25 dex) 
marked by numbers in log scale respectively, and for three values of the 
effective temperature $T_{\rm eff}$ = 33000 K, 40000 K and 45000 K. }
\end{figure*}

%
%
\begin{figure*}
\centerline{\psfig{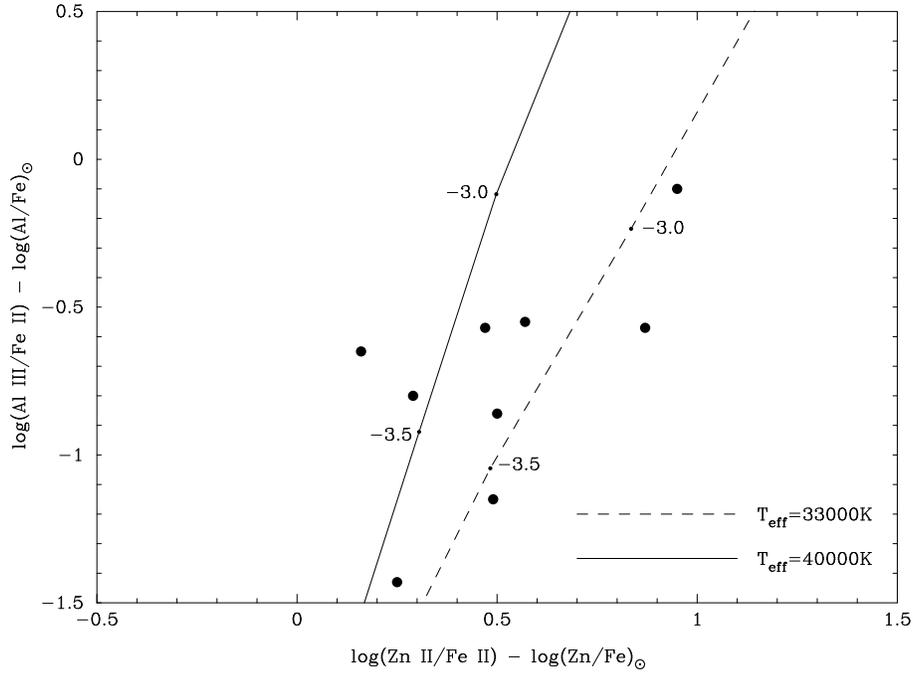}}
\caption{The [Al III/Fe II] vs [Zn II/Fe II] relation. 
The filled circles are
observational points from Prochaska \& Wolfe (1999) and Lu et al. (1996a). 
Dashed and solid lines 
connect theoretical points [$x$(Al$^{+2}$)/$x$(Fe$^+$)], 
[$x$(Zn$^+$)/$x$(Fe$^+$)]
in models with $Z$ = 0.1 $Z_\odot$ for varying values of the ionization 
parameter $U$ 
marked by numbers in log scale respectively, and for two values of the 
effective temperature $T_{\rm eff}$ = 33000 K and 40000 K.}
\end{figure*}

%
%
\begin{figure*}
\centerline{\psfig{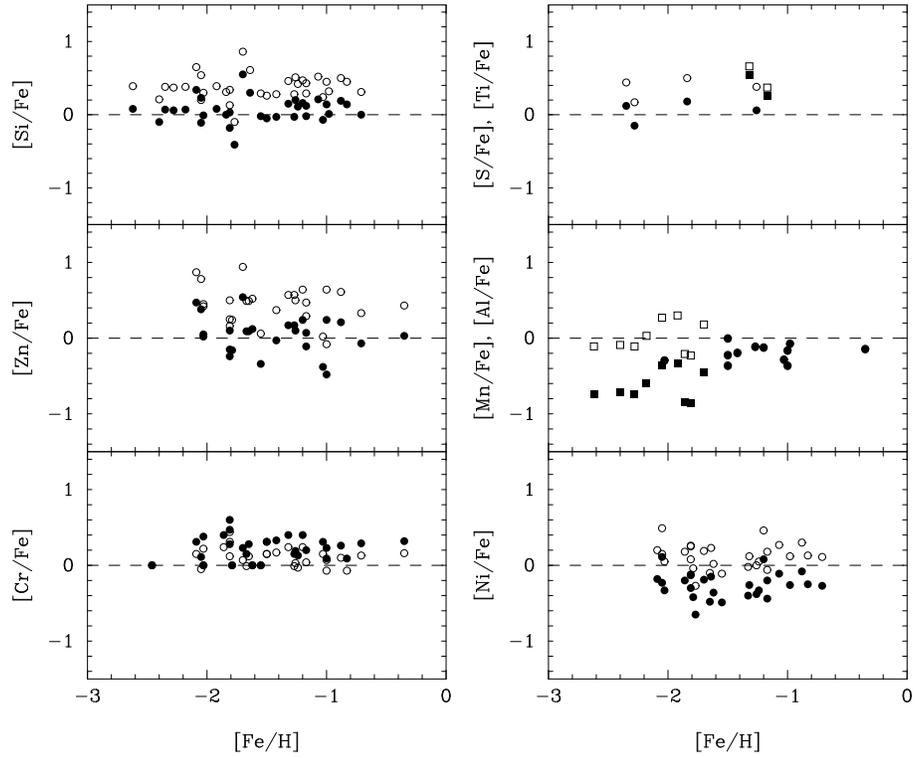}}
\caption{Heavy element abundance ratios for DLAs from
Lu et al.\ (1996), Prochaska \& Wolfe (1999) and Pettini et al.\ (1999).
Open symbols are observed abundance ratios while filled symbols are
abundance ratios corrected for ionization effects. 
Squares are data for [Ti/Fe] and [Al/Fe] while circles are for other
heavy element abundance ratios. 
The ionization-bounded plane-parallel model with ionization 
parameter log $U$ = --3.25 and stellar radiation temperature $T_{\rm eff}$ = 40000 K
is adopted.}
\end{figure*}

 %
%
\begin{figure*}
\centerline{\psfig{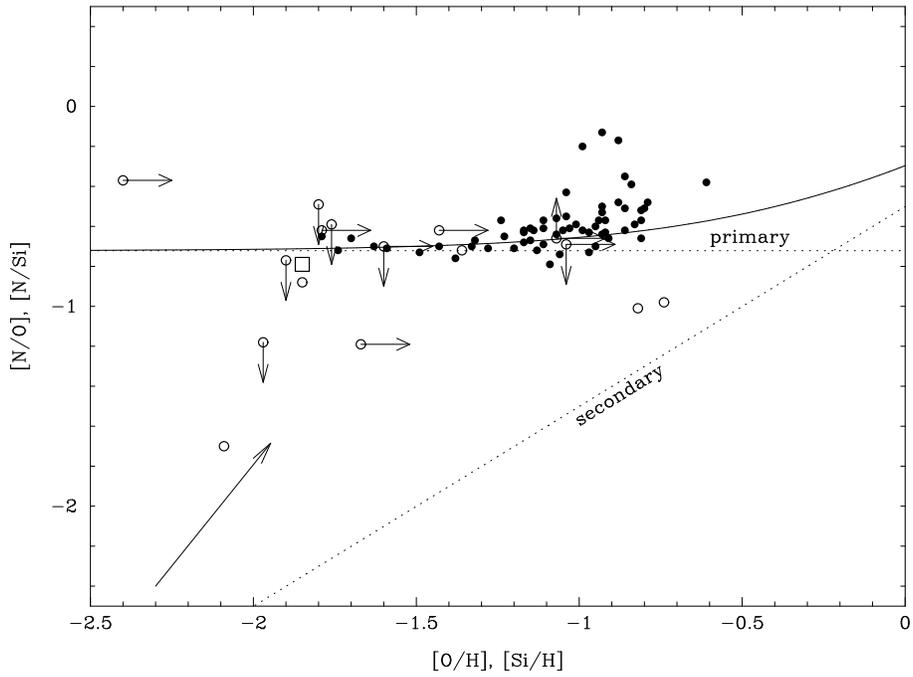}}
\caption{[N/O], [N/Si] vs [O/H], [Si/H] diagram. Open circles
are [N/Si] vs [Si/H] for DLAs from Lu et al.\ (1998) and 
Outram et al.\ (1999), open square are [N/O] vs [O/H] for DLA from Molaro
et al. (2000), 
filled circles are [N/O] vs [O/H] for blue compact dwarf galaxies
from Izotov \& Thuan (1999). Dotted lines are diagrams for primary and
secondary productions of nitrogen, while solid line is dependence [N/O] vs
[O/H] when both primary and secondary productions of nitrogen are taken
into account. Solid line with arrow in the lower left corner shows
the value and direction of abundance
ratio correction for the ionization-bounded plane-parallel model with 
ionization parameter log $U$ = --3.25 and stellar radiation temperature 
$T_{\rm eff}$ = 40000 K.}
\end{figure*}


\begin{references}
\reference {} Anders, S., \& Grevesse, N. 1989, Geochim. Cosmochim. Acta,
53, 197
\reference {} Bergeson, S. D., \& Lawler, J. E. 1993, \apj, 408, 382
\reference {} Centuri\'on, M., Bonifacio, P., Molaro, P., \&
Vladilo, G. 2000, \apj, 536, 540
\reference {} De Mello, D. F., Schaerer, D., Heldmann, J., \& Leitherer, C.
1998, \apj, 507, 199
\reference {} Fedchak, J. A., \& Lawler, J. E. 1999, \apj, 523, 734
\reference {} Ferland, G. J. 1996, HAZY, A Brief Introduction to CLOUDY 90
(University of Kentucky, Department of Physics and Astronomy, Internal Report)
\reference {} Ferland, G. J., Korista, K. T., Verner, D. A., Ferguson, J. W.,
Kingdon, J. B., \& Verner, E. M. 1998, \pasp, 110, 761
\reference {} Forestini, M., \& Charbonnel, C. 1997, \aaps, 123, 241
\reference {} Henry, R. B. C., Edmunds, M. G., \& K\"oppen, J. 2000, \apj, 
541, 660
\reference {} Howk, J. C., \& Sembach, K. R. 1999, \apj, 523, L141
\reference {} Howk, J. C., Savage, B. D., \& Fabian, D. 1999, \apj, 525, 253
\reference {} Hunter, D. A., \& Thronson, H. A., Jr. 1995, \apj, 452, 238
\reference {} Izotov, Y. I., Chaffee, F. H., Foltz, C. B., Green, R. F.,
Guseva, N. G., \& Thuan, T. X. 1999, \apj, 525, 757
\reference {} Izotov, Y. I., \& Thuan, T. X. 1998, \apj, 500, 188
\reference {} ---------. 1999, \apj, 511, 639
\reference {} Jenkins, E. B., Oegerle, W. R., Gry, C., Vallerga, J., 
Sembach, K. R., Shelton, R. L., Ferlet, R., Vidal-Madjar, A., York, D. G.,
Linsky, J. L., Roth, K. C., Dupree, A. K., \& Edelstein, J. 2000, \apj,
538, L81
\reference {} Kunth, D., Lequeux, J., Sargent, W. L. W., \& Viallefond, F.
1994, \aap, 282, 709
\reference {} Kurucz, R. L. 1991, in Proc. Workshop on Precision Photometry:
Astrophysics of the Galaxy, ed. A. C. Davis Philip, A. R. Upgren, \&
K. A. James (Schenectady: Davis), 27

\reference {} Langer, N., Heger, A., Woosley, S. E., \& Herwig, F. 1999,
in Nuclei in the Cosmos, ed. N. Prantzos, in press

\reference {} Lanzetta, K. M., Wolfe, A. M., \& Turnshek, D. A. 1995,
\apjs, 440, 435
\reference {} Lattanzio, J., Forestini, M., \& Charbonnel, C., 1999,
in International  Workshop on ``The changes in abundances in AGB stars'', 
Monteporzio, in press (astro-ph/9912298)
\reference {} Lauroesch, J. T., Truran, J. W., Welty, D. E., \&
York, D. G. 1996, \pasp, 108, 641
\reference {} Le Brun, V., Bergeron, G., Boiss\'e, P., \& Deharveng,
J. M. 1997, \aap, 321, 733
\reference {} Lindner, U., Fritze - von Alvensleben, U., \& Fricke, K. J.
 1999, \aap, 341, 709
\reference {} Lu, L., Sargent, W. L. W., \& Barlow, T. A. 1998, \aj,
115, 55
\reference {} Lu, L., Sargent, W. L. W., Barlow, T. A., Churchill, C. W.,
\& Vogt, S. S. 1996a, \apjs, 107, 475
\reference {} Lu, L., Sargent, W. L. W., Womble, D. S., \& Barlow, T. A. 
1996b, \apj, 457, L1
\reference {} Maeder, A. 1995, in Astrophysical Applications of Stellar
Pulsation, ed. R.S. Stobie, P.A. Whitelock, ASP Conf.\ Series, 83, 1

\reference {} Maeder, A. 1999, in Wolf-Rayet Phenomena in Massive Stars and
Starburst Galaxies, ed. K. A. van der Hucht, G. Koenigsberger, \& P. R. J.
Eenens (Sheridan Books: Michigan), 177
\reference {} Maeder, A., \& Meynet, G. 2000, \araa, in press
\reference {} Maeder, A., Grebel, E. K., \& Mermilliod, J.-C.  1999, \aap, 346, 459 
\reference {} Marigo, P. 1998, PhD thesis, Universita de Padova
\reference {} Marigo, P., Bressan, A., \& Chiosi, C. 1998, \aap, 331, 564
\reference {} Molaro, P., Bonifacio, P., Centuri\'on, M., D'Odorico, S.,
Vladilo, G., Santin, P., \& Di Marcantonio, P. 2000, \apj, 541, 54
\reference {} Outram, P. J., Chaffee, F. H., \& Carswell, R. F. 1999,
\mnras, 310, 289
\reference {} Pettini, M., Ellison, S. L., Steidel, C. C., \& Bowen, D. V.
1999, \apj, 510, 576
\reference {} Pettini, M., Ellison, S. L., Steidel, C. C., 
Shapley, A. E., \& Bowen, D. V. 2000, \apj, 532, 65
\reference {} Pettini, M., King, D. L., Smith, L. J., \& Hunstead, R. W.
1997, \apj, 478, 536
\reference {} Pettini, M., Lipman, K., \& Hunstead, R. W. 1995, \apj,
451, 100
\reference {} Pettini, M., Smith, L. J., Hunstead, R. W., \& King, D. L.
1994, \apj, 426, 79
\reference {} Prantzos, N., \& Boissier, S. 2000, \mnras, 315, 82
\reference {} Prochaska, J., \& Wolfe, A. M. 1996, \apj, 470, 403
\reference {} ---------. 1999, \apjs, 121, 369
\reference {} ---------. 2000, \apj, 533, L5
\reference {} Renzini, A., \& Voli, M. 1981, \aap, 94, 175
\reference {} Ryan, S. G., Norris, J. E., \& Beers, T. C. 1996, \apj, 471, 254
\reference {} Samland, M. 1998, \apj, 496, 155
\reference {} Sembach, K. R., Howk, J. C., Ryans, R. S. I., \& Keenan, F. P.
2000, \apj, 528, 310
\reference {} Stasi\'nska, G., \& Leitherer, C. 1996, \apjs\ 107, 661
\reference {} Thuan, T. X., \& Izotov, Y. I. 1997, \apj, 489, 623
\reference {} Thuan, T. X., Izotov, Y. I., \& Lipovetsky, V. A. 1995, \apj,
445, 108
\reference {} Timmes, F. X., Woosley, S. E., \& Weaver, T. A. 1995,
\apjs, 98, 617
\reference {} Vassiliadis, E., \& Wood, P. R. 1993, \apj, 413, 641
\reference {} Venn, K. A. 1999, \apj, 518, 405 
\reference {} Vidal-Madjar, A. 2000, private communication
\reference {} Vidal-Madjar, A., Kunth, D., Lecavelier des Etangs, A.,
Lequeux, J., Andr\'e, M., BenJaffel, L., Ferlet, R., H\'ebrard, G.,
Howk, J. C., Kruk, J. W., Lemoine, M., Moos, H. W., Roth, K. C., 
Sonneborn, G., \& York, D. G. 2000, \apj, 538, L77
\reference {} Viegas, S. M. 1995, \mnras, 276, 268
\reference {} Vladilo, G. 1998, \apj, 493, 583
\reference {} Wolfe, A. M. 1990, in The Interstellar Medium in Galaxies,
ed. H. A. Thronson \& J. M. Shull (Dordrecht: Kluwer), 387
\reference {} Wolfe, A. M., Lanzetta, K. M., Foltz, C. B., \& Chaffee, F. H.
1995, \apj, 454, 698
\reference {} Wolfe, A. M., \& Prochaska, J. X. 2000, \apj, in press (preprint
astro-ph/0009081)
\reference {} Woosley, S. E., \& Weaver, T. A. 1995, \apjs, 101, 181
\end{references}
\end{document}